# Pore-scale Modelling of Gravity-driven Drainage in Disordered Porous Media

Guanzhe Cui[1], Mingchao Liu[1, 2], Weijing Dai[1] and Yixiang Gan[1*]

[1] School of Civil Engineering, The University of Sydney, NSW 2006, Australia

[2] Department of Engineering Mechanics, CNMM & AML, Tsinghua University, Beijing 100084, China

[*] Corresponding author, email: yixiang.gan@sydney.edu.au

**Abstract:** Multiphase flow through a porous medium involves complex interactions between gravity, wettability and capillarity during drainage process. In contrast to these factors, the effect of pore distribution on liquid retention is less understood. In particular, the quantitative correlation between the fluid displacement and level of disorder has not yet been established. In this work, we employ direct numerical simulation by solving the Navier-Stokes equations and using volume of fluid method to track the liquid-liquid interface during drainage in disordered porous media. The disorder of pore configuration is characterized by an improved index to capture small microstructural perturbation, which is pivotal for fluid displacement in porous media. Then, we focus on the residual volume and morphological characteristics of saturated zones after drainage and compare the effect of disorder under different wettability (i.e., the contact angle) and gravity (characterized by a modified Bond number) conditions. Pore-scale simulations reveal that the highly-disordered porous medium is favourable to improve liquid retention and provide various morphologies of entrapped saturated zones. Furthermore, the disorder index has a positive correlation to the characteristic curve index ($n$) in van Genuchten equation, controlling the shape of the retention characteristic curves. It is expected that the findings will benefit to a broad range of industrial applications involving drainage processes in porous media, e.g., drying, carbon sequestration, and underground water remediation.

**Highlights:**

1. The geometrical features can be quantitatively characterized by the newly defined disorder index $I_v$.

2. With increasing disorder, the formation of local fluid clusters dominates the morphological distribution and promotes higher total residual saturation. Quantitative correlations are derived for describing the fluid displacement with the disorder index.

3. For the small Bond number region (i.e., Bo<1.0), the relationship between the residual saturation and Bond number comply with van Genuchten relation. While for the high Bond number regime, adsorbed liquid dominates for the residual volume.

# 1. INTRODUCTION

Drainage in porous media, a typical multiphase flow problem, plays a significant role in many environmental applications and industrial processes, including carbon capture and storage (CCS), oil and gas extraction, geological hazards, pharmaceutics, and food processes (Parker et al., 1987; Olivella et al., 1994; Bandara et al., 2011; Colombo and Fairweather, 2015; Zhao et al., 2016; Rabbani et al., 2017). As a wetting (e.g., liquid) phase is withdrawn and replaced by another non-wetting (e.g., gas) phase, the flow phenomena and entrapped saturated zone distribution are dominated by complex interactions among different constituents, including gas, liquid and solid, and attracting interests from different research areas related to unsaturated soil mechanics, groundwater remediation, and physics of porous media (Méheust et al., 2002; Islam et al., 2014; Ghanbarian et al., 2015).

The multiphase interactions combining several factors (including intrinsic topological features, gravity, capillary, and wettability) determine macroscopic drainage properties, such as the residual saturation of wetting phase and the temporal/spatial distributions of saturated zones (Yang et al., 1988; Herring et al., 2016; Li et al., 2017). The majority of previous experiments and numerical models of drainage and injection are focusing on the macroscopic parameters, e.g., porosity, permeability, system size and aspect ratio (Succi et al., 1989; Toussaint et al., 2005; Babadagli et al., 2015; Moura et al., 2015; Rognmo et al., 2017; March et al., 2018), in which the spatial configuration, pore connectivity and their influences on liquid retention are ignored (Prat, 1995; Lin et al., 2018; Yekta et al., 2018). However, for most natural and synthetic porous media, disordered microstructures are dominance (Anguy et al., 2001; Woo et al., 2004). These disorder features not only affect the mechanical properties of porous media (Laubie et al., 2017a; Laubie et al., 2017b), but also hinder the fundamental understanding of drainage processes (Holtzman, 2016; Fantinel et al., 2017). Further studies are required to evaluate the degree of microstructural disorder and bridge the microscopic information to macroscopic effective properties of porous media.

Displacement efficiency of two-phase flow in disordered porous media is important in many applications (Dias and Payatakes, 1986). It has been found that the efficiency of immiscible fluids is influenced by the microstructure in those media (Zhao et al., 2016; Hu et al., 2018). For example, low disorder of porous geometry is found to be advantageous for fluid displacement in partially saturated porous media; on the other hand, in terms of liquid mixing and reaction, highly-disordered porous materials is desirable (Holtzman, 2016). These applications indicate that the disordered microstructure and its effects on liquid-liquid displacement need to be quantitatively assessed. However, the quantitative and systematic studies about the effects of disordered microstructure on the multiphase flow problems, particularly on liquid retention during drainage processes, remain relatively scarce.

In the past few years, increasing attention has been focused on the effects of disordered geometry on immiscible two-phase flow (Ferrari et al., 2015; Holtzman, 2016; Rabbani et al., 2017). The ordered and disordered porous media have clear differences in the behaviours and mechanisms of multiphase fluid displacement processes (Pak et al., 2015). With an increase of the disorder degree, the displacement stability is weakened, which results in a lower degree of saturation than that in relatively more homogeneous pore network (Liu et al., 2015). Ferrari et al. (2015) studied numerically the effect of different disordered microstructure controlled by a standard deviation of solid phase size on the unstable invasion structure in Hele-Shaw cells consisting of a large number of cylindrical obstacles. The quantitative assessment on the impact on the disorder of pore distribution keeps relatively unexplored.

In this paper, we employed the direct numerical simulation (DNS) by using the volume of fluid (VOF) method based on Eulerian algorithms to capture the detailed flow dynamics within the porous media, which describes the distribution of two phases as a fluid function with different properties. Then, quasi-two-dimensional (2D) simulations of gravity-driven drainage are performed on a broad range of scenarios and the temporal and spatial information of saturated clusters is exacted and analysed. Our objectives are twofold: firstly, we investigate the reliability of an introduced disorder index to characterize a quantitative relation between the degree of microstructural disorder and retention characteristics; secondly, we demonstrate the effect of gravitational force and wettability on drainage consequence under different disorder degree conditions.

## 2. NUMERICAL MODELING

In this section, we introduce the numerical framework for generating and characterizing the disordered porous media, via an improved disorder index using Voronoi tessellations of the pore space, and modelling the gravity-driven drainage processes employing VOF method in *OpenFOAM*.

## 2.1. Disorder index, $I_v$

To quantitatively investigate the influence of the degree of microstructural disorder in porous media on the drainage process, we explore a meaningful measure of the disorder of the pore space, as well as generating samples with the controlled disorder quantity. Here a model 2D porous medium containing mono-sized circular discs is focused, seen in Fig. 1(a). Random samples can be achieved by deviating the discs from the original regular arrangement, e.g., a hexagonal or square lattice. The porous medium is in form of $N$ disk-shaped solid phases of radius $R$ that is embedded in a rectangular plate of size $L_x \times L_y$ with the periodic boundary condition along the $x$-axis. The centers of discs are moved with a random vector $\vec{R_m}$ ($|\vec{R_m}| \leq \delta$). The disordered porous media contain only non-overlapping discs at a constant global porosity, $\phi$, are constructed and the corresponding degree of disorder can be controlled by varying $\delta$. The movement method of discs allows a larger range of configuration from the completely ordered system, $\delta = 0$, to highly-disordered systems by applying the moving steps iteratively.

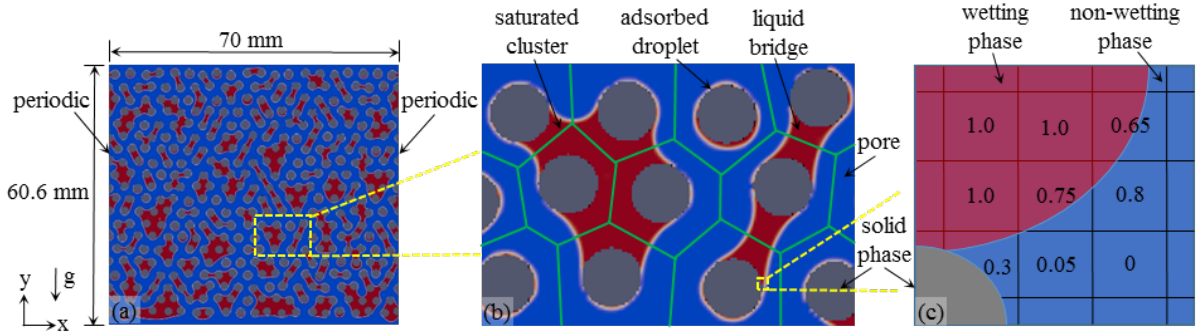

**Fig. 1.** Schematic diagrams of (a) a typical computational domain used for the drainage process containing solid (grey), liquid (red) and gas (blue) phases, (b) Voronoi tessellations on the pore space and three characteristic liquid zones (saturated cluster, adsorbed droplet and liquid bridge), (c) VOF method with the volume fraction of wetting liquid inside each square grid.

Since the porosity $\phi$ as a global parameter is determined by the number and size of discs in the domain, the next high-order parameter is a measurement of the fluctuations of local porosity $\phi_i$, e.g., $I_d = [\langle \phi_i^2 \rangle - \phi^2]^{1/2}$ as the index of disorder proposed by Laubie et al. (2017a). Note that the average operator $\langle \cdot \rangle$ is volumetric average. The local porosity $\phi_i$ is evaluated at every point $i$ in a square background mesh of the domain with a mesh size of $l = (\pi R^2/\phi)^{1/2}$. However, in this paper, we observed that this original definition ignores the small microstructural perturbation which is pivotal for pore fluid flow (discussed later). Thus we propose here a modified index based on Voronoi tessellations of the pore space as the primitive volume, instead of the regular background square mesh, as shown in Fig. 1(b). The local porosity $\phi_i$ can be calculated in a tessellated volume cantered on in the $i$-th pore. The improved disorder parameter ("disorder index"), $I_v$, takes the same format of the original definition, but uses the set $\{\phi_i\}$ based on Voronoi tessellations as

$$I_v = [\langle \phi_i^2 \rangle - \langle \phi_i \rangle^2]^{1/2}. \tag{1}$$

Note here $\phi$ is replaced with the volume average sample porosity, $\langle \phi_i \rangle$. The parameter $I_v$ can change from zero, an ordered geometry, to an approximate maximum value of 0.06 for a highly disordered porous structure.

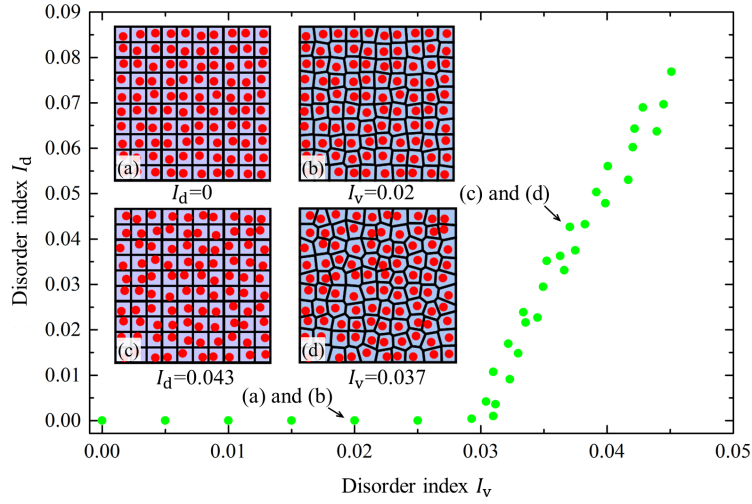

**Fig. 2.** The comparison between disordered indices $I_d$ (original definition) and $I_v$ (improved index). Inserts (a)-(d) indicate the different background meshes used for calculating the local porosity: (a) $I_d=0$ and (b) $I_v=0.02$, (c) $I_d=0.043$ and (d) $I_v=0.037$, where (a-b) and (c-d) are two cases with identical pore structure.

Fig. 2 shows the comparison between two disorder indices, $I_d$ and $I_v$. When the ordered primitive structure is slightly disturbed, e.g., $\delta \leq l/2 - R$, $I_d = 0$, while $I_v$ varies from 0 to 0.03. The reason is that, when the circular discs only randomly move within the square meshes, the local porosity $\phi_i$ equals to mean global porosity $\phi$, as shown in Fig. 2(a)-(d) of the two identical microstructures. This indicates that $I_d$ cannot be used to characterize the small variation in a unit cell (as mentioned as "Randomness Case I" in Laubie et al. (2017b)); on the contrary, the improved definition using Voronoi tessellation is sensitive to such small disturbances. However, this limited movement has a quite significant impact on the drainage processes, which will be demonstrated later in Section 3. As a result, for investigating the pore-scale multiphase flow in this paper, we employed the $I_v$ based on Voronoi tessellations to evaluate the degree of microstructural disorder in porous media. Moreover, the resulting ranges of $I_d$ and $I_v$ also differ from $I_d \in [0,0.1]$ and $I_v \in [0,0.06]$, respectively.

## 2.2. Volume of Fluid

In traditional computational fluid dynamics, during the procedure of solving the Navier-Stokes equations, it is extremely difficult and computationally expensive to solve moving interfacial boundary condition problems (Horgue et al., 2013). However, the VOF method employs treatments of two immiscible and isothermal, incompressible liquid phases as a single-fluid phase with different properties including density and viscosity. It utilizes an additional interfacial force replacing the jump condition at the liquid-liquid interface (Owkes and Desjardins, 2014). In VOF method, the process of drainage in porous media can be described as a wetting phase ("defending fluid") replaced by a non-wetting phase ("invading fluid") (Ferrari and Lunati 2013).

The spatial distribution of two immiscible fluid phases is described as the phase indicator for fluids, shown in Fig.1 (c), as

$$\alpha = \begin{cases} 0 & \text{in the non-wetting (nw) fluid,} \\ 1 & \text{in the wetting (w) fluid.} \end{cases} \quad (2)$$

According to the dynamic law of fluid phase, the immiscible fluids can be expressed by the conversation of mass and momentum as a set of Navier-Stokes equations:

$$\nabla \cdot \boldsymbol{u} = 0, \quad (3)$$

$$\frac{\partial \rho \boldsymbol{u}}{\partial t} + \nabla \cdot (\rho \boldsymbol{u}\boldsymbol{u}) = -\nabla p + \nabla \cdot (2\mu \boldsymbol{E}) + \boldsymbol{f}_s, \quad (4)$$

where $\boldsymbol{u}$ is the fluid velocity, $p$ is the pressure, $\boldsymbol{E} = \frac{1}{2}(\nabla u + \nabla u^T)$ is the strain rate tensor, $\rho$ and $\mu$ are the density and viscosity considering multiple fluid phases, respectively, as:

$$\rho(\boldsymbol{x}) = \alpha \rho_w + (1-\alpha)\rho_{nw}, \quad (5)$$

$$\mu(\boldsymbol{x}) = \alpha \mu_w + (1-\alpha)\mu_{nw}, \quad (6)$$

In Eq. (4), the last term represents the effect of Laplace pressure at the interface and is defined as:

$$\boldsymbol{f}_s = 2\sigma k \boldsymbol{n} \delta_T, \quad (7)$$

where $\sigma$ is the interfacial tension between the two fluids, $\delta_T$ is a unit function that is 0 in the whole domain except at the interface, $\boldsymbol{n}$ is the normal to the interface, and $k$ is the mean curvature at the interface:

$$k = \frac{1}{2} \nabla \cdot \boldsymbol{n} = \frac{1}{2} \nabla \cdot \left( \frac{\nabla \alpha}{\|\nabla \alpha\|} \right). \tag{8}$$

In this work, the no-slip condition is adopted, i.e., the vectorial component of velocity is zero on the solid boundary. To reproduce the wetting behaviour at the triple line region, the boundary condition in VOF equals to Young's law as defined as (Ferrari and Lunati, 2013):

$$\boldsymbol{n}|_{\delta_\mathrm{r}} = \boldsymbol{t_s}\sin\theta + \boldsymbol{n_s}\cos\theta, \tag{10}$$

where $\boldsymbol{n}$ is the normal to two fluids interface at the solid surface, $\boldsymbol{t_s}$ is the unit tangent toward the wetting phase, $\boldsymbol{n_s}$ is the unit normal pointing into the solid, and $\theta$ is the contact angle. When the above equations are used to compute the whole-domain velocity field and describe the evolution of the fluid-fluid displacement, the advection equation should be included as

$$\frac{\partial \alpha}{\partial t} + \nabla \cdot (\alpha \boldsymbol{u}) + \nabla \cdot (\boldsymbol{u}_\mathrm{r} \alpha (1 - \alpha)) = 0, \tag{11}$$

where $\boldsymbol{u}_\mathrm{r}$ ($\boldsymbol{u}_\mathrm{r} = \boldsymbol{u}_1 - \boldsymbol{u}_2$) is the relative velocity between the two fluids. Since *OpenFOAM* use the surface compression method to implement the VOF method, the second term on the left hand side is an artificial compression term applied to prevent the numerical diffusion which smears the sharp interface (Boyce et al., 2016).

## 2.3. Simulation setup and materials parameters

The numerical simulations are implemented in an open-source CFD platform, *OpenFOAM*, with a multiphase flow solver *interFoam*. This solver employs Finite Volume (FV) schemes for the discretization of partial differential equations to calculate multiple immiscible fluids phases. Inside the pore space, the classical Navier-Stokes equations are solved by VOF method tracking the temporal and spatial evolution of fluid-fluid interface. A mesh generator, *snappyHexMesh*, is used to automatically generate complex meshes of hexahedral and split-hexahedral cells from triangulated surface geometry in the domain. In order to construct a quasi-2D model, the out-of-plane thickness of the sample is particularly set to be much smaller than the in-plane edges in x and y axis. The initial background hexagonal mesh is iteratively refined to better reproduce the cylindrical surface with the meshes.

To investigate the morphological characteristics of entrapped wetting fluid, including absorbed liquid, liquid bridge and saturated cluster, as shown in Fig. 1(b), we simulated drainage under atmospheric condition in a rectangular domain (Fig. 1(a)) with inlet on top, outlet on bottom and periodic boundary condition in the horizontal direction. The geometrical parameters of computational models are shown in Table 1. The dynamics of fluid-fluid displacement, morphology and distribution of wetted zone during gravity-driven drainage depend on the properties of wetting and non-wetting phases including density, viscosity and surface tension, which are selected based on the parameters of water and air (see Table 1).

**Table 1.** Geometrical parameters of the quasi-two-dimensional porous media and properties of two immiscible fluids used in numerical simulations.

| Property | Value |
|---|---|
| Radius of discs, $R$ (mm) | 1 |
| Contact angle, $\theta$ (°) | 0°~150° |
| Viscosity of wetting phase (m$^2$/s) | 5.93×10$^{-7}$ |
| Viscosity ratio, $\mu_w/\mu_{nw}$ (-) | 0.38 |
| Density of wetting phase, $\rho_w$ (kg/m$^3$) | 997 |
| Density of non-wetting phase, $\rho_n$ (kg/m$^3$) | 11.69 |
| Density difference, $\Delta\rho$ (kg/m$^3$) | 985.31 |
| Void ratio, e (-) | 2.65 |
| Gravity acceleration, g (m/s$^2$) | 9.81 |
| Surface tension, $\gamma$ (N/m) | 0.064 |

**Table 2.** A sensitivity study on the domain size based on the same discretization ($2R/\Delta=27$).

|  | Small | Medium | Large |
|---|---|---|---|
| Dimensions (mm$^2$) | 35×52.32 | 70×96.13 | 140×193.25 |
| Number of discs (-) | 100 | 400 | 1600 |
| Number of meshes (-) | 825,547 | 2,975,769 | 11,990,042 |
| Porosity (-) | 0.704 | 0.705 | 0.704 |
| Degree of residual saturation | 17.73% | 16.90% | 14.94% |

**Table 3.** A sensitivity study on discretization based on the medium-sized domain.

|  | Discretisation, $2R/\Delta$ | | | |
|---|---|---|---|---|
|  | 16 | 27 | 32 | 41 |
| Number of meshes (-) | 1,080,298 | 2.975,769 | 5,268,631 | 6,662,617 |
| Porosity (-) | 0.704 | 0.705 | 0.705 | 0.708 |
| Degree of residual saturation | 15.86% | 16.90% | 17.11% | 16.93% |

Before the detailed simulation of drainage processes, we have performed sensitivity studies on the mesh discretization and domain size to identify the cost-efficient mesh density and representative model size. A compromise between the level of discretization and size of computational domains has to be found. To this end, we tested several cases including three computational domains and four levels of discretization, defined by the ratio of disc diameter to typical mesh cell size (i.e., $2R/\Delta=16, 27, 32, 41$). The benchmark is defined in terms of the residual saturation at the end of gravitational drainage process, and all cases are conducted under the same porosity and pore structure, with $\theta=30°$ and Bond number of 0.36. Based on these results (shown in Table 2 and Table 3), we decided to select the case with medium domain size of 70×96.13 mm$^2$ with 400 solid discs and discretisation of $2R/\Delta=27$ (meshes), which ensures that differences with the results obtained with the finest mesh discretization and largest domain size are within 3%.

## 3. RESULTS ANALYSIS

By employing the proposed numerical model in the last section, we can investigate the gravity-driven drainage processes. More specifically, in this section, the effects of degree of disorder, gravity and wettability on the residual volume and morphology distribution of wetting phase will be discussed.

### 3.1. Effects of disorder index, $I_v$

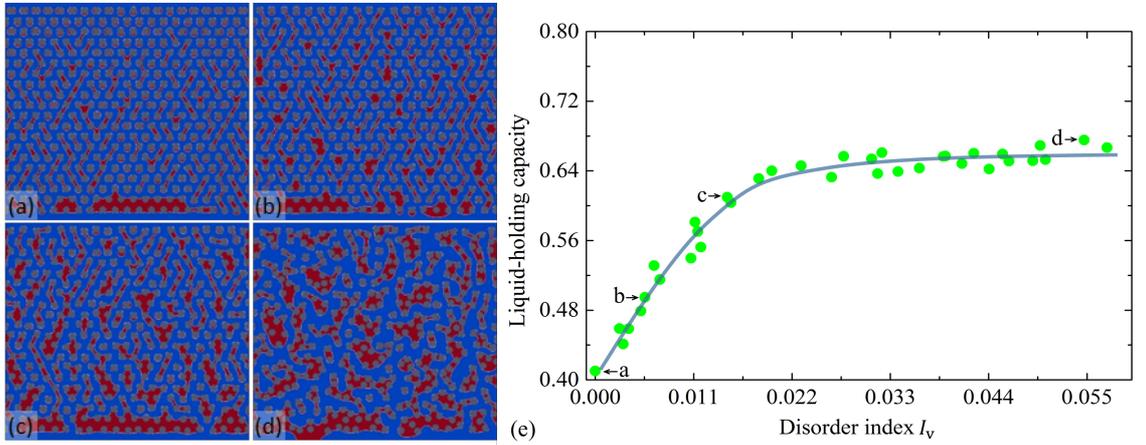

**Fig. 3.** Simulation results of four different samples with increasing disorder index: (a) $I_v$=0, (b) $I_v$= 0.0056, (c) $I_v$ =0.015 and (d) $I_v$=0.055. (e) Liquid-holding capacity (the residual volume of wetting phase normalized by the solid volume as a function of disorder index $I_v$. The scale shows the range of liquid-holding capacity varying from 0.41 to 0.68.

We first demonstrate the reliability of disorder index $I_v$ by characterizing the effect of disorder degree of pore topology on the liquid-holding capacity defined as the residual volume of wetting phase normalized by the solid volume. Note that, in order to exclude upper and lower boundary effects in the calculation, the volume of residual liquid within the top one row and bottom two rows of discs are removed during image processing. Fig. 3(a)-(d) show the spatial distributions of liquid zones (i.e., the entrapped saturated zones) in four samples with different pore topologies after drainage, which correspond to disorder index: $I_v$=0 (ordered distribution), $I_v$=0.0056, $I_v$=0.015 and $I_v$=0.055 (highly-disordered topology), respectively.

The morphological characteristics of entrapped saturated zones are significantly different in the four samples. Such variations in the morphology of the entrapped saturated zones at pore-scale will be analysed in detail in Section 3.2. In the macroscopic scale, the numerical results demonstrate that the degree of disorder has a monotonously increasing relationship with the volume of entrapped saturated zones, as shown in Fig. 3(e), in which the solid line is for a guiding purpose. It is obvious that $I_v$ varying in the range from 0 to 0.02 has a more significant impact on the residual volume than the rest range. Absorbed droplets are rapidly connected into liquid bridges and even saturated clusters with the increasing degree of disorder within this range, which are more efficient to enhance the liquid-holding capacity than the other $I_v$ range that mainly contains clusters. Note that $I_d$ in the corresponding range keeps nearly zero when $I_v$ vary from 0 to 0.02 (see Figure 2), failing to capture this transition region. These results indicate that the parameter $I_v$ is effective to quantitatively evaluate the influence of the topological disorder on the liquid holding capacity. In addition, the increase of the disorder index is unfavourable for drainage efficiency in porous media.

### 3.2. Distribution of residual saturation cluster

In this section, the study aims to find a relationship between the disorder degree and the morphological characteristics of entrapped saturated zones after drainage. The spatial information of the saturated clusters is extracted by image processing. In statistics, the area of liquid bridges range from 0.8 mm² to 4 mm², and other zones of smaller or larger areas correspond to absorbed droplets and saturated clusters, respectively, with typical examples shown in Fig. 1(b). In the post-processing, the extremely small liquid volume with a size smaller than one-tenth of a solid disc area ($\pi$ mm²) volume is neglected.

The probability density distribution of saturated zones volume after drainage processes are plotted in Fig. 4. The distributions of zone area using a uniform bin-width of 0.4 mm² are extracted from four samples with different disorder degree (i.e., $I_v$=0, $I_v$=0.0056, $I_v$=0.015 and $I_v$=0.055), corresponding to the pore topologies in Fig. 3(a)-(d), respectively. Fig. 4(a) shows that, when the solid discs are assigned regularly ($I_v$=0), the maximum probability of cluster area is possessed by the bin from 0.4 mm² to 0.8 mm², which is mostly consisted of the absorbed droplets. In Fig. 4(b) and Fig. 4(c), the span of the size distribution of saturated zones becomes wider with increasing disorder of pore structure. Specifically, the three kinds of wetting phase morphology (absorbed droplet, liquid bridge and saturated cluster) simultaneously exist at the final drainage stages. When disorder index arrives the approximately maximum value ($I_v$=0.055), the saturated zones in the highly-disordered porous media have the maximum cluster area reaching 40 mm² as well as the minimum number of absorbed droplets and liquid bridges (see Fig. 4(d)). These results demonstrate that the increase of disorder contributes to the aggregation of solid discs, which is advantageous for the merging of absorbed droplets into liquid bridges and further into residual saturated clusters.

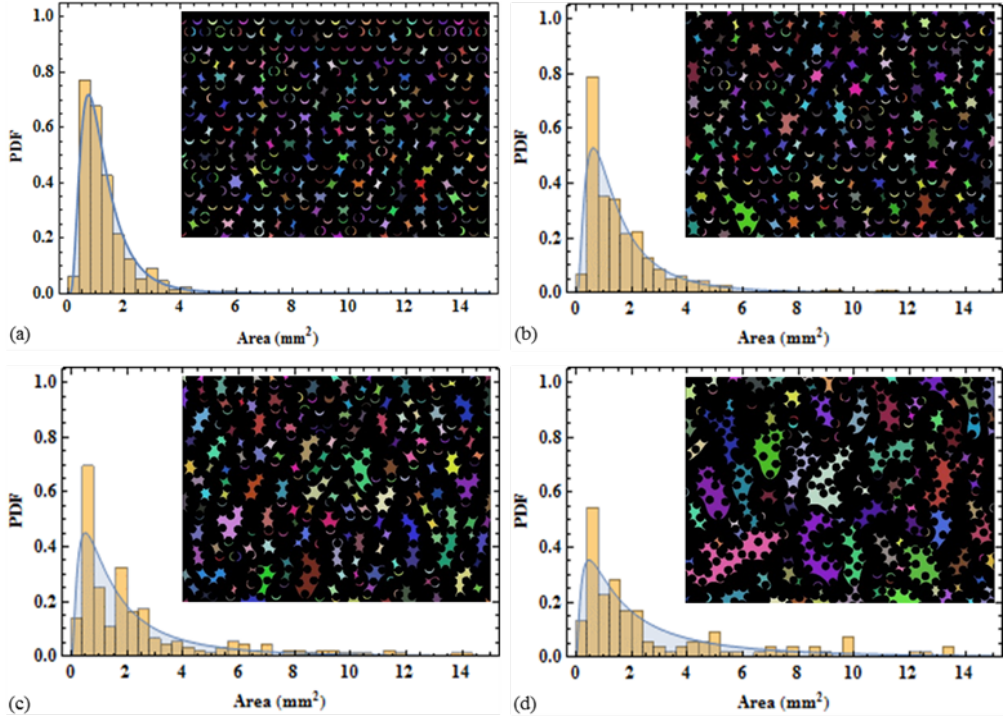

**Fig. 4.** Probability density function (PDF) and lognormal distribution of saturated zones area and maps of morphological characteristics with the four disorder parameters: (a) $I_v$=0, $\mu$=-0.061, $\sigma$=0.59; (b) $I_v$=0.0056, $\mu$=0.24, $\sigma$=0.85; (c) $I_v$=0.015, $\mu$=0.38, $\sigma$=1.04; (d) $I_v$=0.055, $\mu$=0.66, $\sigma$=1.20. Note that the different colours in the inserted morphological character map represent the disconnected wetting phase zones.

The dataset can be characterized by employing the lognormal distribution function, which has been employed in interpreting experimentally observed saturation clusters (Li et al. 2017) as

$$f(X|\mu,\sigma^2) = \frac{1}{\sqrt{(2\pi\sigma^2)}X} exp\left[-\frac{(lnX-\mu)^2}{2\sigma^2}\right]. \tag{12}$$

According to the maximum likelihood estimation method, two determinative parameters, i.e., the scale parameter ($\mu$) and shape parameter ($\sigma$), can be obtained by fitting the probability density distribution of liquid zone areas. It is clearly seen from Fig. 4(a)-(d) that the mean $\mu$ and variance $\sigma$ have a positive relationship with the disorder index, indicating the increase of the mean and variation of residual liquid areas of high degree of disorder by enhancing the formation of saturated clusters. The reason is that a large $I_v$ improves the solid phase aggregation, individual liquid bridges and absorbed droplets merging and eventually leading to the formation of large saturated zones in those dense regions. This further illustrates why the liquid-holding capacity is gradually enhanced with the increase of disorder degree and how the disorder of porous media statistically varies the total volume and morphological characteristics of residual wetting phase after the gravity-driven drainage process.

### 3.3. Liquid retention under varying Bond number

During the gravity-driven drainage process, the interplay between the gravity and capillary forces has a significant effect on the fluid displacement behaviours. In order to quantify the relative magnitude of these two forces, a dimensionless parameter, i.e., the Bond number, is introduced in the previous work (Méheust et al., 2002; Moebius and Or, 2014) as

$$B_o = \frac{\Delta \rho g a^2}{\gamma}, \qquad (13)$$

where $a$ is the characteristic length, $\Delta \rho$ density difference between wetting and non-wetting phases, $g$ the gravity acceleration, and $\gamma$ the surface tension. The characteristic length is usually defined as the throat size (Blunt and Scher, 1995) or the radius of pore (Birovljev et al., 1991; Løvoll et al., 2005). However, because the disorder degree introduces non-uniform solid disc separations, the throat size is not easy to quantify. Here we adopt the uniform radius of solid discs as the characteristic length. In order to consider the variation due to the void ratio, we introduce the volumetric ratio between the pore and solid phase into the traditional form Eq. (13) and obtain the modified Bond number:

$$B_o' = \frac{\Delta \rho g R^2}{\gamma} \cdot e, \qquad (14)$$

where $R$ is the radius of the solid disc and $e = V_P / V_S$ is the void ratio, in which $V_P$ and $V_S$ are the total volume of pore and solid phase, respectively.

To demonstrate the validity of the modified Bond number, the parameters including the density difference, the gravity acceleration, the radius of solid disc, the void ratio are individually adjusted with the surface tension to keep the $B_o'$ the same at 0.37. Fig. 5 demonstrates that the degree of saturation is similar when the $B_o'$ is kept the same for given contact angle, across a range of wettability. In the following analyses, we will focus on the combined effects among the modified Bond number, disorder and wettability.

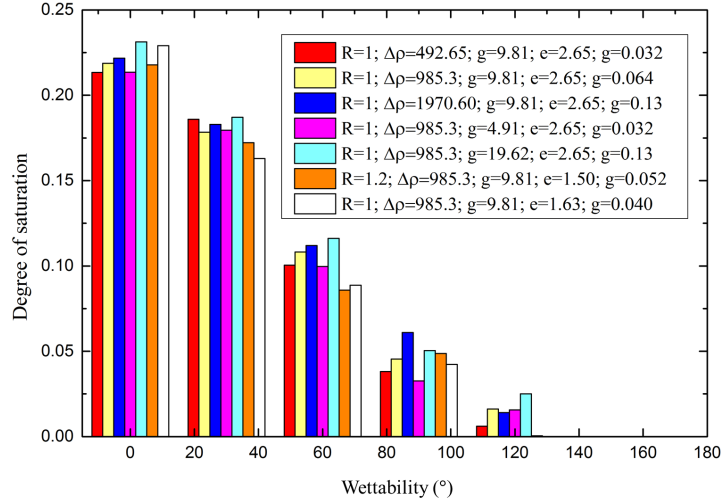

**Fig. 5.** Degree of residual saturation as a function of wettabilities at $B_o'=0.37$ with a different combination of density difference ($\Delta\rho$, kg/m$^3$), gravity acceleration ($g$, m/s$^2$), disc radius ($R$, mm), void ratio ($e$) and surface tension ($\gamma$, N/m).

To elucidate the combined effect of disorder degree and the modified Bond number, $B_o'$, simulations are performed for the four disordered samples as shown in Fig. 3(a)-(d) with a broad range of $B_o'$ from 0.011 to 7.70. The liquid-holding capacity of the disordered samples is plotted in Fig. 6 as a function of $B_o'$. The results demonstrate that, based on the influence of disorder, the drainage process can be divided into two categories, the capillary regime and adsorbed regime. The $B_o'$ threshold separating these two categories is approximately 1.0, beyond which the gravitational force overtakes capillary force in drainage process, resulting in the disappearance of saturation clusters and capillary bridges.

When $B_o'$ is less than the threshold, capillarity dominates the fluid-fluid displacement. Based on comparison between the ordered and the highly disordered medium (see Fig. 6(a) and 6(b)), higher disorder can dramatically improves the liquid-holding capacity of porous media with the same $B_o'$; in contrast to the only small absorbed droplets in the ordered system, high disorder leads to large entrapped saturated zones, which demonstrates that disorder degree has a significant effect on the capillary interaction during drainage. In addition, it can be clearly seen that there is a morphological transformation from saturated zones (Fig. 6(a)) to liquid bridge (Fig. 6(c)) with the increasing gravitational force.

In the second regime, i.e. $B_o' > 1.0$, the gravitational force has overcome the capillary counterpart. Fig. 6(d) indicates that only the absorbed droplets exist on the surface of solid discs instead of the saturated patch and liquid bridges between discs in Fig 6(b) and 6(c), respectively. The absorbed volume gradually declines with the increase of Bond number. Since the absorbed liquid phase is only proportional to the total surface area (length in 2D) of the solid phase, there is no observed difference for ordered and disordered pore spaces, exhibiting similar relationship as shown in Fig. 6.

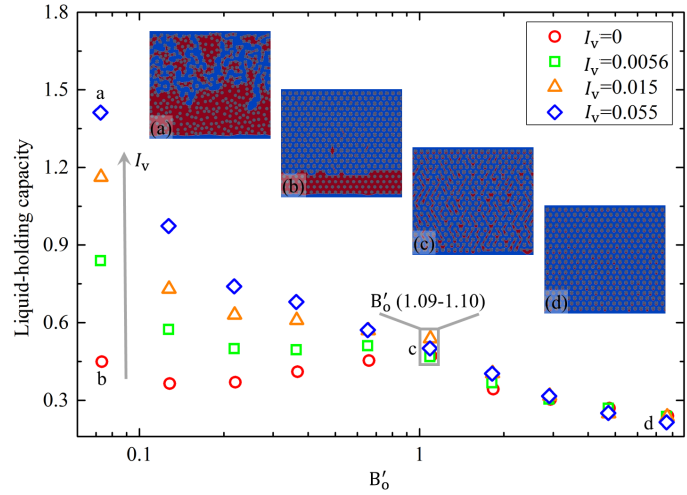

**Fig. 6.** Liquid-holding capacity as a function of Bond number with four different disorder indices and maps of wetted cluster distribution at contact angle $\theta=30°$: (a) $I_v=0.055$, $Bo'=0.073$, (b) $I_v=0$, $Bo'=0.073$, (c) $I_v=0$, $Bo'=1.10$, (d) $I_v=0.027$, $Bo'=7.65$.

The van Genuchten equation originally describes the relationship between the degree of saturation and matric suction in the soil-water characteristics curves (SWCCs) (Zhou et al., 2016). As an analogue, here the residual volume is governed by the interaction between the capillary and gravitational forces (characterized by Bond number). Thus, we replaced the suction term in the original format with the Bond number:

$$S_r = \frac{1-S_r^a}{[1+(B_o'/\alpha)^n]^{1-1/n}} + S_r^a, \quad (15)$$

where $S_r$ is the degree of saturation, $S_r^a$ the adsorbed saturation, $\alpha$ the scaling parameter positively proportional to the air-entry value, and $n$ is the characteristic curve index controlling the slope of SWCC (Fredlund et al., 2012). Fig. 7 shows that the liquid retention curve can be quantitatively predicted by the van Genuchten equation (i.e., Eq. (15)). Furthermore, we conclude that with the same porosity and solid disc diameter, the increase of disorder index $I_v$ decreases the curve index *n*. During the curve fitting process, we used $\alpha=0.055$ and $S_r^a=0.20$, since the former was found to be insensitive to $I_v$ and latter can be measured at $B_o^{'}=1$. Thus, the potential relationship between the disorder of pore space and liquid characteristic curves can be evaluated quantitatively by the application of $I_v$ under the framework of unsaturated soil mechanics.

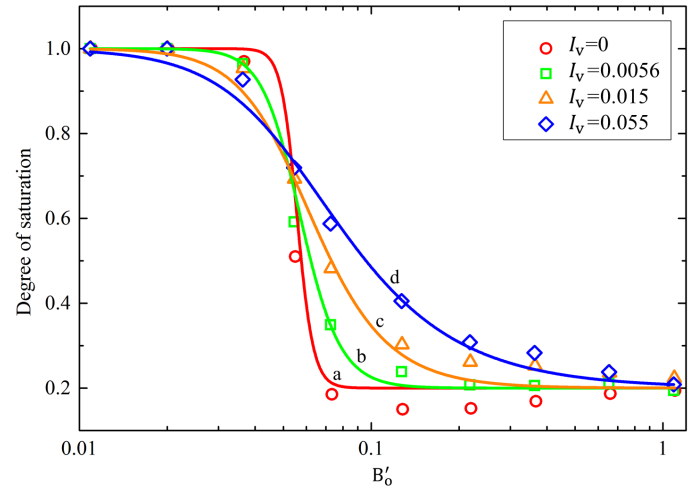

**Fig. 7.** Degree of residual saturation as a function Bond number fitted by van Genuchten equation (with $\alpha=0.055$, $S_r^a=0.20$): (a) $I_v=0$, $n=16.07$, (b) $I_v=0.0056$, $n=6.80$, (c) $I_v=0.015$, $n=3.75$, (d) $I_v=0.055$, $n=2.54$.

### 3.4. Surface wettability

It is well known that the wettability of porous media can dramatically change the liquid-liquid displacement behavior during both drainage and imbibition processes (Zhao et al., 2016), but whether the disorder alters such influence of wettability has yet been explored. To this end, a series of simulations were performed on the disordered porous samples with the wettability of solid surface varying over a broad range from 0° to 150° at the same Bond number of 0.37. The results are plotted in Fig. 8. It is clearly seen that, with the increase of disorder index $I_v$, the volume of liquid retention is gradually elevated at the same wettabilities, which indicates the influence of the wettability on retention can be controlled by adjusting the disorder index (see Fig. 8(a) and 8(b)). Meanwhile, as for varying wettability, it is relatively more efficient to alter the residual volume when the disorder index is within $I_v \in (0-0.02)$, than the larger degree regime, $I_v \in (0.02-0.06)$. Due to the disorder-induced reinforcement of liquid-holding capacity, the maximum and minimum relative difference, compared to the ordered configuration at the same wettability, are 108% at the hydrophilic ($\theta=60°$) and 716% at hydrophobic ($\theta=150°$) samples, respectively.

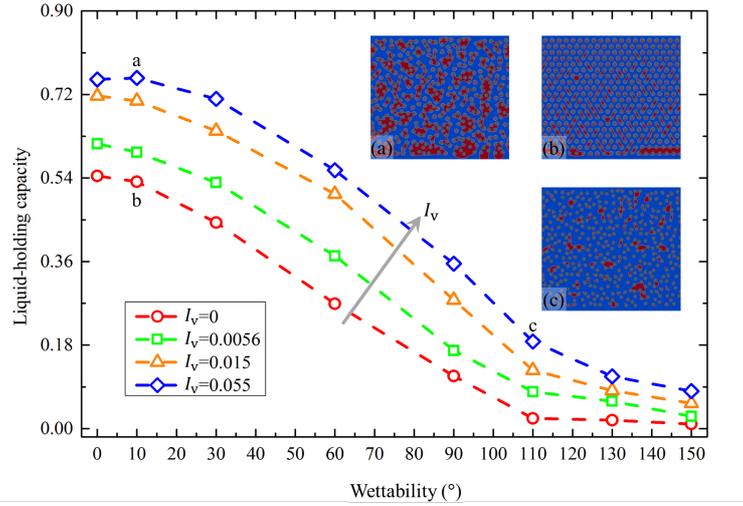

**Figure. 8.** Liquid-holding capacity as a function of wettability for the different disorder degree and morphological patterns of drainage results with different wettability conditions as inserts: (a) $I_v$=0.055, $\theta$=10°; (b) $I_v$=0, $\theta$=10°; and (c) $I_v$=0.055, $\theta$=110°.

In addition, when the surface wettability of porous media varies from hydrophilic to hydrophobic, wetting phase begins to flow out more and the liquid-holding capacity drastically decreases until approximately the inflection point (i.e., $\theta$~110°). During this transition, in terms of high disorder degree, the morphological characteristics of final drainage clearly transforms from a variety of liquid clusters to only entrapped droplets between the circular solid phase (see Fig. 8(c)). Such entrapment can be noticeably strengthened in high disorder samples and eventually keeps the residual volume of wetting phase nonzero. When the contact angle is within 0° to 110°, the effect of wettability on drainage efficiency is more obvious.

## 4. DISUCUSSIONS

Our numerical results have shown that the disorder parameter $I_v$ is reliable to quantitatively characterize the influence of disorder degree on liquid retention in porous media after the gravity-driven drainage process. The residual volume has a monotonously increasing relationship with the disorder index under the same porosity and radius of solid discs. It demonstrates that highly-disordered media is disadvantageous for drainage efficiency, however, the efficiency can be more sensitively controlled within the range of disorder index $I_v$ (0-0.02) than the larger range (0.02-0.06).

For representing the morphological characteristics, we employed the lognormal function to fit the probability density distribution of saturated areas. The scale parameters $\mu$ and shape parameter $\sigma$ are positively correlated with disorder index $I_v$, which indicates that enhancing disorder is advantageous for the formation of large fluid clusters. Due to the reinforcement of solid phase aggregation, a large quantity of absorbed droplets and liquid bridges can be gradually connected into liquid clusters.

In addition, a modified Bond number, including the void ratio, is introduced. Based on the magnitude of gravitational and capillary forces, the drainage process is divided into two categories and the transition point of is approximately $B_o'=1.0$. When the gravitational force dominates the drainage process ($B_o'>1.0$), only absorbed droplets exist on the solid surface without the presence of saturated clusters and liquid bridges, resulting in almost the same liquid-holding capacity for different disorder degree; on the other hand, when capillary force is larger than gravitational force ($B_o'<1.0$), the relationship between the degree of saturation and Bond number can be well represented by van Genuchten equation, in which disorder index $I_v$ has a positive relationship with the characteristic curve index $n$.

Finally, the study of fluid drainage characteristics with various surface wettability demonstrates that the effect of microstructural disorder on residual volume is clearly weakened with the increase of hydrophobicity. In addition, when the surface wettability vary from hydrophilic to hydrophobic, morphological characteristics of highly-disordered topology have an obvious transformation from various saturated zones to only entrapped droplets.

In summary, this paper provides a fresh perspective into the significant applications of two-phase flows, in which displacement efficiency and morphological feature of wetting phase are crucial. For example, decreasing disorder degree can control the drainage processes, desirable for improving the displacement efficiency of water during drying of food and soil aeration processes, and accelerating the recovery of contaminants and $CO_2$ geosequestration. On the other hand, increasing disorder enlarges the interfacial area, which is advantageous for the subsurface remediation, fluid-gas mixing, and reaction enhancement in microfluidics.

## 5. CONCLUSIONS

Based on Voronoi tessellations of the pore space, the disorder index $I_v$ has been introduced to describe the drainage characteristics in disordered porous media. With the help of this index, we studied the gravity-driven drainage with different pore configurations exhibiting a broad range of disorder degree. The results demonstrate that $I_v$ is able to quantitatively characterize the influence of the disorder of porous media on the residual volume and morphological characteristics of residual saturated zones. Among other factors (capillarity, gravity, wettability, porosity and characteristic size), this study highlights the correlations between the disorder index to the following drainage behaviour: (1) residual saturation, (2) morphological features and (3) statistical distribution of residual wetting phase. The results presented in this paper contribute to fundamentally understand the physical principles behind multiphase flow in disordered porous media and to improve applications related to control and optimize the drainage processes.

## ACKNOWLEDGEMENTS

This work was financially supported by Australian Research Council (Projects DP170102886) and The University of Sydney SOAR Fellowship.